\documentclass[11pt]{article}
\usepackage{moriond,epsfig}

\def\Journal#1#2#3#4{{#1} {\bf #2}, #3 (#4)}

\def\PLB{{\em Phys. Lett.}  B}

\def\be{\begin{equation}}
\def\ee{\end{equation}}
\def\bea{\begin{eqnarray}}
\def\eea{\end{eqnarray}}

\def\Jpsi{J/\psi}
\def\NPA{{\em Nucl. Phys.} A}
\def\EPJC{{\em Eur. Phys. J.} C}
\begin{document}
\vspace*{4cm}
\title{RESULTS ON $\psi'$ PRODUCTION IN NUCLEUS--NUCLEUS COLLISIONS AT
CERN-SPS}
\author{ M.\ SITTA$^{1}$ (for the NA50 Collaboration)\\
B.~Alessandro$^{10}$,
C.~Alexa$^{3}$,
R.~Arnaldi$^{10}$,
M.~Atayan$^{12}$,
S.~Beol\`e$^{10}$,
V.~Boldea$^{3}$,
P.~Bordalo$^{6,a}$,
G.~Borges$^{6}$,
C.~Castanier$^{2}$,
J.~Castor$^{2}$,
B.~Chaurand$^{9}$,
B.~Cheynis$^{11}$,
E.~Chiavassa$^{10}$,
C.~Cical\`o$^{4}$,
M.P.~Comets$^{8}$,
S.~Constantinescu$^{3}$,
P.~Cortese$^{1}$,
A.~De~Falco$^{4}$,
N.~De~Marco$^{10}$,
G.~Dellacasa$^{1}$,
A.~Devaux$^{2}$,
S.~Dita$^{3}$,
J.~Fargeix$^{2}$,
P.~Force$^{2}$,
M.~Gallio$^{10}$,
C.~Gerschel$^{8}$,
P.~Giubellino$^{10}$,
M.B.~Golubeva$^{7}$,
A.A.~Grigorian$^{12}$,
S.~Grigoryan$^{12}$,
F.F.~Guber$^{7}$,
A.~Guichard$^{11}$,
H.~Gulkanyan$^{12}$,
M.~Idzik$^{10,b}$,
D.~Jouan$^{8}$,
T.L.~Karavitcheva$^{7}$,
L.~Kluberg$^{9}$,
A.B.~Kurepin$^{7}$,
Y.~Le~Bornec$^{8}$,
C.~Louren\c co$^{5}$,
M.~Mac~Cormick$^{8}$,
A.~Marzari-Chiesa$^{10}$,
M.~Masera$^{10}$,
A.~Masoni$^{4}$,
M.~Monteno$^{10}$,
A.~Musso$^{10}$,
P.~Petiau$^{9}$,
A.~Piccotti$^{10}$,
J.R.~Pizzi$^{11}$,
F.~Prino$^{10}$,
G.~Puddu$^{4}$,
C.~Quintans$^{6}$,
L.~Ramello$^{1}$,
S.~Ramos$^{6,a}$,
L.~Riccati$^{10}$,
H.~Santos$^{6}$,
P.~Saturnini$^{2}$,
E.~Scomparin$^{10}$,
S.~Serci$^{4}$,
R.~Shahoyan$^{6,c}$,
F.~Sigaudo$^{10}$,
M.~Sitta$^{1}$,
P.~Sonderegger$^{5,a}$,
X.~Tarrago$^{8}$,
N.S.~Topilskaya$^{7}$,
G.L.~Usai$^{4}$,
E.~Vercellin$^{10}$,
L.~Villatte$^{8}$,
N.~Willis$^{8}$,
T.~Wu$^{8}$}

\address{
$^{~1}$ Universit\'a del Piemonte Orientale, Alessandria and INFN-Torino, Italy
$^{~2}$ LPC, Univ. Blaise Pascal and CNRS-IN2P3, Aubi\`ere, France
$^{~3}$ IFA, Bucharest, Romania
$^{~4}$ Universit\`a di Cagliari/INFN, Cagliari, Italy
$^{~5}$ CERN, Geneva, Switzerland
$^{~6}$ LIP, Lisbon, Portugal
$^{~7}$ INR, Moscow, Russia
$^{~8}$ IPN, Univ. de Paris-Sud and CNRS-IN2P3, Orsay, France
$^{~9}$ LPNHE, Ecole Polytechnique and CNRS-IN2P3, Palaiseau, France
$^{10}$ Universit\`a di Torino/INFN, Torino, Italy
$^{11}$ IPN, Univ. Claude Bernard Lyon-I and CNRS-IN2P3, Villeurbanne, France
$^{12}$ YerPhI, Yerevan, Armenia
\\
a) also at IST, Universidade T\'ecnica de Lisboa, Lisbon, Portugal
b) also at Faculty of Physics and Nuclear Techniques, Academy of Mining 
and Metallurgy, Cracow, Poland
c) on leave of absence of YerPhI, Yerevan, Armenia
}

\maketitle\abstracts{
We report results on $\psi'$ production as a function of centrality 
as measured by experiment NA50, at the CERN/SPS, in Pb-Pb collisions at 
an incident energy of 158~GeV per nucleon.  
} 

\section{Introduction}
Charmonia production in proton--proton, proton--nucleus and nucleus--nucleus
collisions has been addressed since the mid 80s by the NA38 and the NA50
experiments. The yield of $c {\bar c}$ bound states in such interactions can be
affected by absorption in nuclear matter or, potentially, by interactions with
hadrons produced in the collision. Moreover, according to non-perturbative QCD
predictions, ultrarelativistic nucleus-nucleus collisions could lead to a phase
transition of normal nuclear matter  to a quark-gluon deconfined state (QGP)
inducing, through the so-called {\it Debye colour screening} \cite{matsui}, a
specific suppression of the production of the J/$\psi$ bound state. The NA50
experiment has already measured a significant abnormal $\Jpsi$ suppression in
Pb-Pb collisions \cite{abreu00,ramello03}. Within this framework, the
production of $\psi'$ has also been measured although, as a more loosely bound
state, it is expected to be much less specific of a phase transition.  

\section{Data selection and analysis method}
The NA50 apparatus \cite{abreu97}, an upgrade of the previous NA38 detector, is
designed to detect and measure muon pairs in an air gap toroidal magnet muon
spectrometer. The centrality of the collision can be estimated from a
Multiplicity Detector (MD) measuring the charged multiplicity of the
interaction, an Electromagnetic Calorimeter (EM) measuring the neutral
transverse energy $E_T$\/, or a Zero-Degree Calorimeter (ZDC) measuring the
beam ion spectators energy $E_{ZDC}$\/.

The Pb-Pb data used for the $\psi'$ analysis reported here were collected in
1998 and 2000, with a Pb ion beam impinging on a single Pb target, placed in
air in 1998 and in vacuum in the 2000 setup. Dimuons are selected in the
rapidity range $2.92 \leq y_{lab} < 3.92$ ($0 \leq y_{CM} < 1$) and with a
Collins--Soper angle $| \cos\vartheta_{CS} | < 0.5$, leading to an acceptance
of $\sim 14\%$ in the dimuon invariant mass region of interest. After rejecting
upstream interactions and the residual pile-ups by means of dedicated
detectors, on-target interactions are identified requiring the appropriate
correlation between hits in the MD and applying proper track quality cuts (see
\cite{ramello03}). A summary of the analyzed sample is given in
Table~\ref{tab:data}.
\begin{table}[t]
\caption{Summary of the data samples presented here.\label{tab:data}}
\vspace{0.4cm}
\begin{center}
\begin{tabular}{|c|c|c|c|c|c|}
\hline
{\it Data taking}&{\it Target}&{\it Number of}&
{\it Beam intensity}&{\it Number}&{\it Number}\\
{\it period}&{\it thickness}&{\it subtargets}&
{\it (ions/burst)}&{\it of $\Jpsi$}&{\it of $\psi'$}\\
\hline
1998 & $7 \% \lambda_I$ & 1 (in air) &
$5.5 \cdot 10^7$ & \phantom{1}49000 & 380 \\
2000 & $9.5 \% \lambda_I$ & 1 (in vacuum) &
$7.0 \cdot 10^7$ & 129000 & 905 \\
\hline
\end{tabular}
\end{center}
\end{table}
The events are analyzed as a function of the interaction centrality. In each
centrality class the opposite-sign dimuon mass spectrum results from four
physical components ($\Jpsi$, $\psi'$, Drell-Yan and semi-leptonic $D\bar{D}$
decays) and the combinatorial background. The physical components are estimated
through a fit where the shape of each component is obtained by a MonteCarlo
simulation which includes the apparatus description and operation and the data
reconstruction (Drell-Yan and open charm are generated at leading order using
PYTHIA \cite{pythia}). The combinatorial background, mostly due to $\pi$ and
$K$ decays, is estimated from the like-sign muon pairs in the real data
according to $N_{BG} = 2\sqrt{N^{++}N^{--}}$.
\begin{figure}[h]
\vskip 2.5cm
\psfig{figure=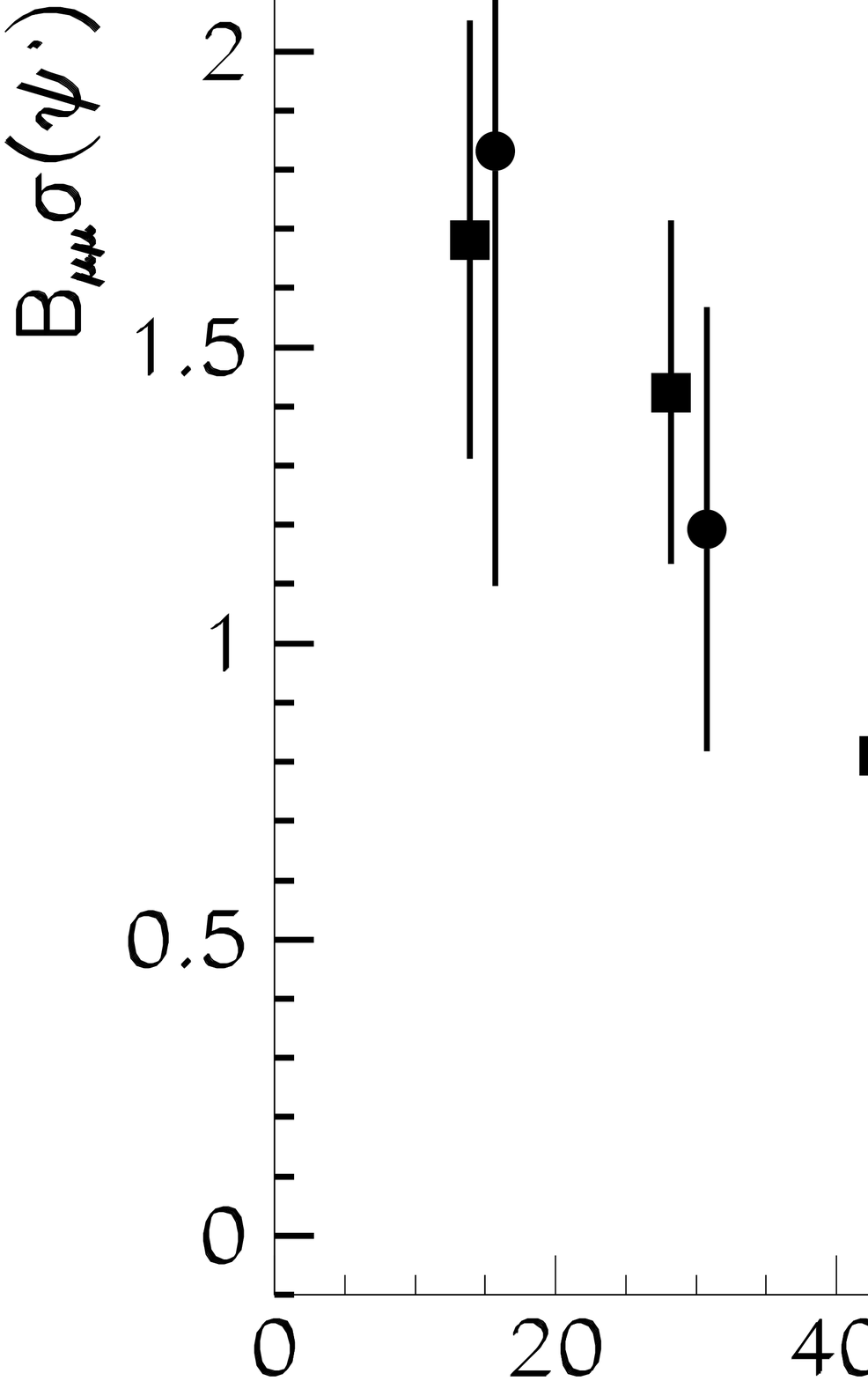,height=2.9cm}
\psfig{figure=psip_Jpsi.ps,height=2.9cm}
\caption{$B_{\mu \mu}'\sigma(\psi')/\sigma(DY)$ (left) and $B_{\mu \mu}'\sigma
(\psi')/B_{\mu \mu}\sigma(J/\psi)$ (right) as a function of $E_T$ for the Pb-Pb
1998 and 2000 data samples.
\label{fig:yields}}
\end{figure}

\section{$\psi'$ production in Pb-Pb collisions}
NA38 and NA50 have verified that Drell-Yan is proportional to the number of
elementary nucleon-nucleon interactions from p-p up to Pb-Pb
collisions \cite{bordalo02}. The $\psi'$ production rate is therefore
normalized to Drell-Yan, which, moreover, cancels out most of the systematic
effects. The fit to the dimuon mass spectrum in each centrality region leads to
the $\psi'$ and the Drell-Yan yields (the latter in the mass interval $4.2 -
7.0$~Gev/$c^2$\/), and hence to the ratio $B_{\mu \mu}'\sigma
(\psi')/\sigma(DY)$, proportional to the $\psi'$ production cross-section per
elementary collision. This ratio is displayed in Fig.\ \ref{fig:yields} along
with the ratio $B_{\mu \mu}'\sigma(\psi')/B_{\mu \mu}\sigma(J/\psi)$\/, as a
function of $E_T$. Besides the good compatibility between the two samples, it
shows that the $\psi'$ production rate is increasingly suppressed with
centrality, and that the $\psi'$ is more suppressed than $\Jpsi$ with
increasing centrality.

\section{Comparison with lighter systems}
\begin{figure}[h]
\vskip 2.3cm
\psfig{figure=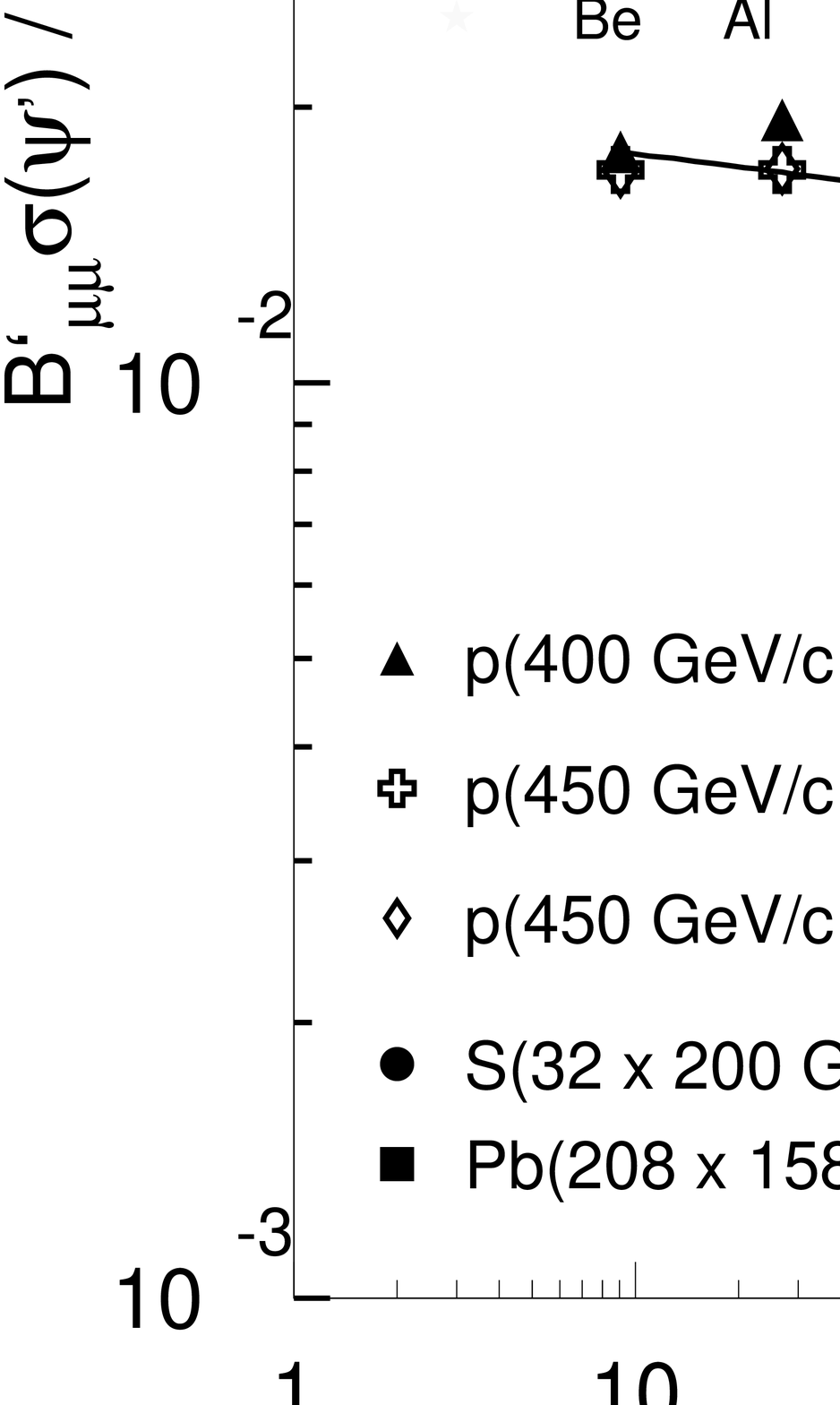,height=2.9cm}
\psfig{figure=psip_L.ps,height=2.9cm}
\caption{$B_{\mu \mu}'\sigma(\psi')/B_{\mu \mu}\sigma(J/\psi)$ as a function of
$A \times B$ (left) and $B_{\mu \mu}'\sigma(\psi')/\sigma(DY)$ as a function of
$L$ (right).
\label{fig:lighter}}
\end{figure}

The $\psi'$ suppression in Pb-Pb collisions can be compared with the results
obtained from both proton-induced reactions by the same NA50 experiment (at
$400$ and $450$~GeV/$c$ incident p momentum) and S-U reactions by the NA38
experiment (at $200$A GeV/$c$ incident Sulphur momentum). The relative
suppression of the two charmonia states as a function of the product of the
projectile and target atomic mass numbers $A$ and $B$ is shown in Fig.\
\ref{fig:lighter} (left). The ratio $B_{\mu \mu}'\sigma(\psi')/B_{\mu \mu}
\sigma(J/\psi)$ can be parametrized with a power law $A^{\Delta \alpha}$, where
$\alpha$ accounts for all nuclear effects. The measured negative value of
$\Delta \alpha$ indicates that the $\psi'$ is more suppressed than the $\Jpsi$
already in p-A collisions \cite{alessan,borges}. Moreover the $\psi'$ appears
to be even more suppressed in ion-ion interactions.
Fig.\ \ref{fig:lighter} (right) shows the ratio $B_{\mu \mu}'\sigma(\psi')/
\sigma(DY)$ as a function of $L$, the mean path crossed by the $c {\bar c}$
pair in the nuclear matter ($L$ is estimated using the Glauber model). The
behaviour of the production rate is clearly different in p-A and A-B
interactions. Using the parametrization ${\rm e}^{- < \rho L > \sigma_{abs}}$, 
where $\rho$ is the nuclear density, we obtain $\sigma_{abs} = 7.4 \pm 1.4$~mb 
in p-A collisions and the much higher value $\sigma_{abs} = 21.6 \pm 2.5$~mb in
ion-ion collisions. Furthermore, the $\psi'$ suppression is the same in S-U and
Pb-Pb interactions as a function of $L$.

\section{Comparison between $\Jpsi$ and $\psi'$ productions}
\begin{figure}[h]
\vskip 2.6cm
\psfig{figure=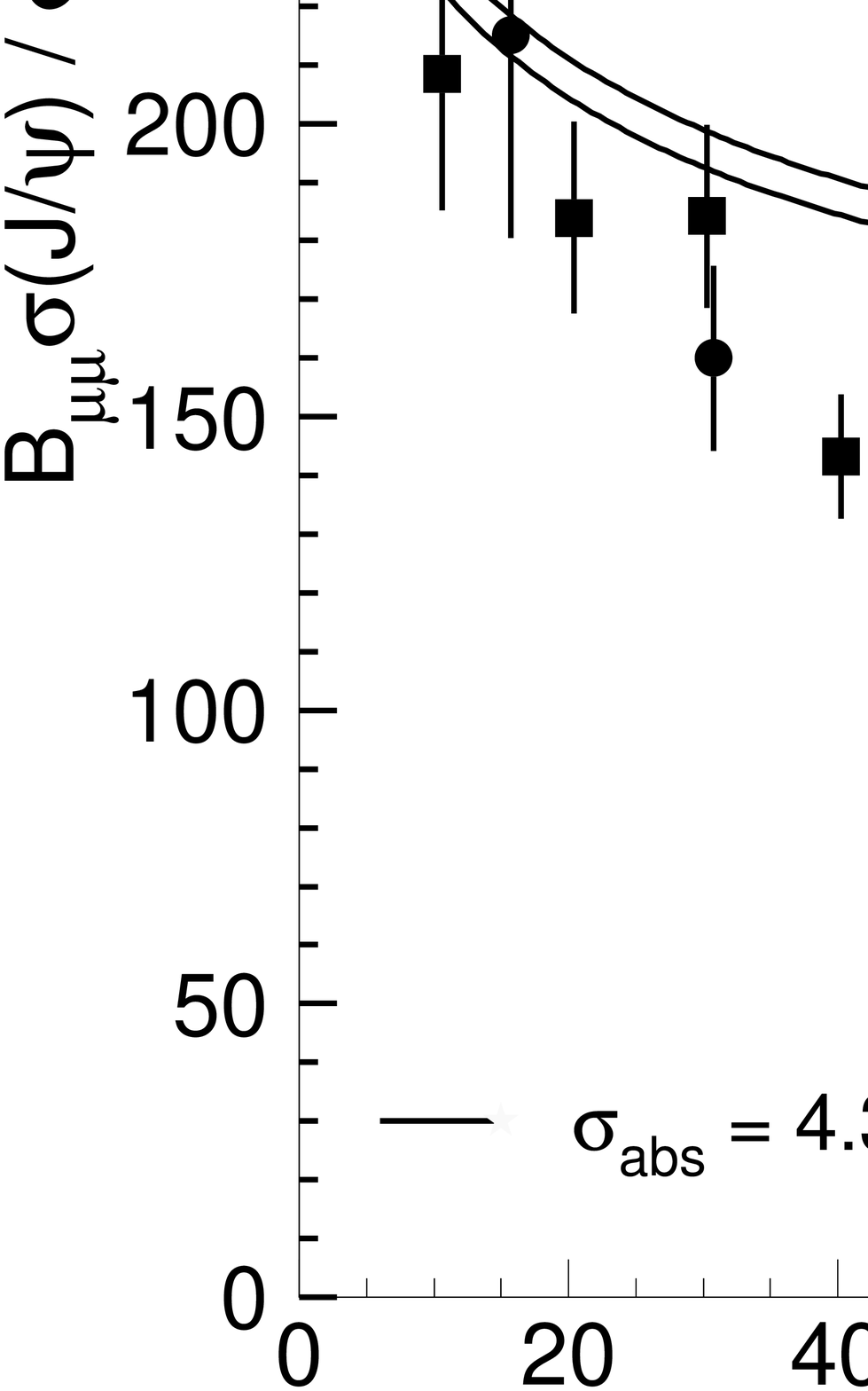,height=3.1cm}
\psfig{figure=meas_expc.ps,height=3.1cm}
\caption{$B_{\mu \mu}\sigma(\Jpsi)/\sigma(DY)$ as a function of $E_T$ (left)
and the ratio ``measured value''/''expected value'' for the relative yields
$B_{\mu \mu}\sigma(\Jpsi)/\sigma(DY)$ and $B_{\mu \mu}'\sigma(\psi')/\sigma
(DY)$ as a function of $L$ (right).
\label{fig:final}}
\end{figure}

Fig.\ \ref{fig:final} (left) presents updated results on the ratio $B_{\mu \mu}
\sigma(\Jpsi)/\sigma(DY)$ as a function of $E_T$ in Pb-Pb
collisions \cite{borges}. The superimposed absorption curve is obtained from
p-A and S-U data \cite{borges}, and the band accounts for the Drell-Yan energy
rescaling uncertainties. While for peripheral collisions the data points follow
the standard absorption curve, they  later depart from this normal behaviour
with no clear flattening at high $E_T$, a pattern which qualitatively agrees
with the theoretically predicted anomalous $\Jpsi$ suppression.
Fig.\ \ref{fig:final}(right) shows the ratio of the measured charmonia yields
to the corresponding ``expected'' values as a function of $L$ for various
interacting systems. The expected values are computed only from normal
absorption in nuclear matter, using a full Glauber calculation with
$\sigma_{abs} = 4.3 \pm 0.3$ mb for the $\Jpsi$, as extracted from a
simultaneous fit to p-A and S-U data, and with $\sigma_{abs} = 7.9 \pm 0.6$ mb
for the $\psi'$, as extracted from p-A data only (since this absorption value
is incompatible with the one obtained in S-U data). The plot shows that in A-B
collisions the $\psi'$ departs from the expected absorption curve for lower $L$
values with respect to the $\Jpsi$.

\section{Conclusions}
In Pb-Pb collisions the $\psi'$ is suppressed by a factor 7 with respect to
Drell-Yan and by a factor $2.5$ with respect to $\Jpsi$ between peripheral and
central collisions. In comparison with lighter systems the $\psi'$ is much more
suppressed in nucleus-nucleus than in proton-nucleus interactions, and the
suppression pattern is the same in S-U and Pb-Pb as a function of $L$. The
anomalous suppression for the $\psi'$ sets in for lower $L$ values than for the
$\Jpsi$.

\end{document}